\documentstyle[12pt,A4wide]{article}
\begin{document}
\rightline{June the 3rd, 1992}
{\begin{center} {\Large Energy Spectrum of Anyons\\
\vspace{0.3cm}
                        in a Magnetic Field} \\

\vspace{2cm}
{\large Fabrizio Illuminati\footnote{Bitnet
address: illuminati@padova.infn.it}}  \\
\vspace{0.3cm}
{\it Dipartimento di Fisica ``Galileo Galilei'', Universit{\`a} \\
di Padova, Via F.Marzolo 8, Padova 35131, Italia.} \end{center}}

\vspace{2cm}
{\begin{center} \large \bf Abstract \end{center}}

For the many-anyon system in external magnetic field,
we derive the energy spectrum as an
exact solution of the quantum eigenvalue problem with
particular topological constraints. Our results agree with
the numerical spectra recently obtained for the 3- and
the 4-anyon systems.

\vspace{0.4cm}
PACS numbers: 03.65.Ge, 03.65.Sq, 05.30.-d

\vfill
DFPD 92/TH/42 \hfill June 1992

\newpage
{\large \bf 1. Introduction}

\vspace{0.4cm}

The possibility of arbitrary statistics continously
interpolating between the Bose and the Fermi case in
space dimensions less than three was suggested some
years ago in a beautiful and pioneering paper [1] by
Leinaas and Myrheim. A physical model of particles with
arbitrary, or fractional, statistics was put forward
by Wilczek [2], who baptized such objects as ``anyons".
Anyon dynamics has lately received much attention (for
comprehensive recent reviews and guides to the literature
see ref. [3]-[5]), as it is thought to be a good
candidate for the explanation of such important
collective phenomena in condensed matter physics as the
fractional quantum Hall effect and high temperature
superconductivity. To this end, it is then crucial
to investigate the dynamics of the many-anyon systems.

The symmetry group of the multi-anyon state vectors is
not the group of permutations, it is the braid group.
Due to this fact, even in the non-interacting case the
many-anyon problem is not separable, and the exact
solution has been obtained (by Leinaas and Myrheim
in their original paper [1])
only for the case of $N=2$ non-interacting anyons in a
harmonic well. For the three-anyon problem Wu has obtained
a special set of exact solutions [6] which have been
recently generalized to the arbitrary $N$-anyon system in
a magnetic field by Chou [7] and by Dunne {\em et al.} [8].
In fact, the problem of $N$ anyons moving in an external
harmonic potential can be easily translated in that of
$N$ charged anyons moving in a constant magnetic field.
All the enrgy levels obtained from the special exact
solutions [6]-[8] exhibit a strict linear dependence
on the statistical parameter $\alpha$, as in the exactly
solved 2-anyon problem [1].

Recently, very important experiments have been performed
by Sporre, Verbaarschot and Zahed, who have solved
numerically the Schroedinger eigenvalue problem for
the $N=3$ and the $N=4$ anyon systems [9],[10].
The energy spectrum so obtained presents some levels
with a slight non-linear dependence on $\alpha$, and some
with a strictly linear dependence on $\alpha$. The latter
include Wu's exact eigenvalues.

In coincidence with the numerical investigations,
Illuminati, Ravndal and Ruud have introduced a
semi-classical approximation to determine the energy
spectrum of the
3-anyon system [11]. The energy levels are again linear
in $\alpha$ and their slopes agree with the numerical ones.
We have quite recently generalized the semi-classical
quantization scheme to the general $N$-anyon system in
external harmonic potential [12]. The semi-classical
solution is obtained in connection with an exact limit
of separability of the general many-anyon Hamiltonian, and
it is shown to reproduce and explain all the main features
of the numerical results also in the 4-anyon problem.

In this paper we wish to address the many-anyon problem
in external magnetic field in the framework of the
exactly solvable model introduced in [12]. We will
solve the eigenvalue problem from topological
considerations specific of the classical trajectories
of charged particles in a magnetic field. The
energy spectrum so obtained reproduces not only
the slopes (as in the case of the harmonic potential)
but also the actual values of the numerical energy levels.

\vspace{1cm}

{\large \bf 2. Classical Trajectories in a Magnetic Field}

\vspace{0.4cm}

Consider the motion of a classical non-relativistic
particle with mass $m$ and charge $e$ in a constant
magnetic field of strength $B$. Assume the field
directed along the $z$-axis and the particle
constrained to the $xy$-plane (e.g. when the
velocity along the field is zero). The particle
then moves in cyclotron orbits with angular
velocity $\omega = eB/m$. Choosing the
symmetric gauge ${\bf A} = \frac{1}{2} {\bf B
\times r}$, the lagrangian equation of motion
for the particle has the general solution

\begin{equation}
{\bf r}(t) = r_{0}[(\cos\omega t){\bf i} -
(\sin\omega t){\bf j}] + {\bf R} ,
\end{equation}

\noindent where $r_{0}$ is the radius of the
cyclotron orbit, ${\bf R}$ is the position
vector of the center of the orbit, and
${\bf i}, {\bf j}$ are the unit vectors along
the $x$- and the $y$-axis respectively.
The energy $E$ and magnitude $L$ of the
angular momentum of the particle are

\begin{eqnarray}
E & = & \frac{1}{2}m{\omega}^{2}r_{0}^{2} , \\
L & = & \frac{1}{2}m\omega(R^{2} - r_{0}^{2}) .
\end{eqnarray}

We see from eq.(3) that the classical orbits
fall into two different topological classes,
whether they enclose the origin of coordinates
or not. When the angular momentum is positive
the orbits do not enclose the origin of
coordinates. Negative values of $L$ correspond
instead to orbits enclosing the origin.
This feature of the classical orbits plays
an important role in the semi-classical
analysis of the many-anyon problem.

\vspace{1cm}

{\large \bf 3. The two-anyon system}

\vspace{0.4cm}

In the symmetric gauge, the Lagrangian for $N$ anyons
in a magnetic field is

\begin{equation}
L = \frac{1}{2}\sum_{i=1}^{N}(m{\bf \dot{r}}_{i}^{2} +
e{\bf \dot{r}}_{i}\cdot{\bf B} \times {\bf r}_{i}) -
\alpha\hbar\sum_{i<j}{\dot{\phi}}_{ij} ,
\end{equation}

\noindent where ${\bf r}_{i} = (x_{i}, y_{i})$, and the
azimuthal angle $\phi_{ij}$ is defined by

\begin{equation}
\phi_{ij} = \arctan\frac{y_{j} - y_{i}}{x_{j} - x_{i}}.
\end{equation}

The statistical parameter $\alpha$ can be restricted to
to take positive values in the interval $[0, 1]$. Bose
satistics is recovered for $\alpha = 0$ and Fermi statistics
for $\alpha = 1$. We could choose to reverse the sign
of the topological term in the Lagrangian, but then
$\alpha$ should vary in the interval $[-1, 0]$ to
obtain the correct statistical influence of the
winding numbers on the energy eigenvalue problem.
The situation here is inverted respect to the
case of the harmonic oscillator: there the topological
term enters the Lagrangian with a positive relative
sign, if $\alpha$ is chosen to be positive.

We now specialize to the case $N = 2$. Introducing
polar coordinates $(\rho, \phi)$, the
separation of the center of mass motion is
immediate, and the effect of statistics is all
contained in the Lagrangian for the relative motion

\begin{equation}
L_{rel} = \frac{m}{2}({\dot{\rho}}^{2} +
{\rho}^{2}{\dot{\phi}}^{2}) + (\frac{e}{2}
B{\rho}^{2} - \alpha\hbar)\dot{\phi} .
\end{equation}

The corresponding Hamiltonian is

\begin{equation}
H = \frac{1}{2m}(p_{\rho}^{2} + \frac{1}{{\rho}^{2}}(p_{\phi}
+ \alpha\hbar)^{2} - eB(p_{\phi} + \alpha\hbar) +
\frac{1}{4}e^{2}B^{2}{\rho}^{2}) .
\end{equation}

Apart from the terms containing $\alpha$, eq. (7) is exactly
the Hamiltonian for an ordinary particle in a magnetic
field. The eigenvalue problem is readily solved, in the same
way as for the two-dimensional
oscillator with statistical interaction originally
treated in [1]. In terms of the Laguerre polynomials the
eigenfunctions of $H$ are

\begin{equation}
{\psi}_{n,m} = e^{-\frac{1}{2}{\xi}^{2}}L_{n}^{|m
+ \alpha|}({\xi}^{2})e^{im\phi}{\xi}^{|m + \alpha|} ,
\end{equation}

\noindent where $\xi = \rho\sqrt{m\omega / 2\hbar}$.
The corresponding energy eigenvalues (modified
Landau levels) are

\begin{equation}
E_{n,m} = (2n + 1 + |m + \alpha| - m -
\alpha)\frac{\hbar\omega}{2} ;
\end{equation}

\noindent the radial quantum number $n$ is a positive
integer, while the angular quantum number must be
an even integer (positive or negative) if we model
the two anyons as bosons with statistical interaction,
and thus require the wave functions to be symmetric.
The spectrum is drawn in Fig.1.
{}From eq. (9) it follows that the energy is
independent of
the angular momentum and of $\alpha$ when $m > 0$. From
the discussion of section (2) it corresponds to the classical
relative orbit not encircling the origin of coordinates, i.e.
to the two absolute classical orbits being separated and not
winding around each other (see Fig.2a). In turn, a zero
winding number implies the corresponding quantum result
that the energy does not depend on $\alpha$. A second class
of levels depending linearly on $\alpha$ with slope $-2$ is
obtained when $m < 0$. The corresponding classical relative
orbit encircles the origin, i.e. the orbit of one particle
winds around the orbit of the other (Fig.2b), generating a
non-zero winding number which, in turn, makes the corresponding
energy eigenvlues depend on $\alpha$.

\vspace{1cm}

{\large \bf 4. Separation of the many-anyon problem}

\vspace{0.4cm}

A semi-classical formula for the energy spectrum of $N$
harmonically bound anyons was derived in [12]. There
the semi-classical result emerges
as the exact solution of the quantum problem in a
suitable limit of separability of the many-body
Hamiltonian; it is the limit of non-crossing winding
orbits. The same argument applies to the $N$-anyon
system in a magnetic field. Here we are motivated by
the fact that the exact solution of the two-anyon
problem corresponds to classically
separated cyclotron orbits. We then assume that this
property holds in general for $N > 2$. Introducing
$N - 1$ Jacobi relative coordinates $\{ {\mbox{
\boldmath $\rho$}}_{k} \}$ in polar form

\begin{eqnarray}
{\rho}_{k} & = & \frac{1}{\sqrt{k(k - 1)}}(r_{1} +
r_{2} + \cdots + r_{k - 1} - (k - 1)r_{k}) , \nonumber \\
{\phi}_{k} & = & \arctan\frac{y_{1} + y_{2} +
\cdots + y_{k} - ky_{k + 1}}{x_{1} + x_{2} +
\cdots + x_{k} - kx_{k + 1}} , \nonumber \\
k & = & 1, 2, \ldots, N - 1 ,
\end{eqnarray}

\noindent the separation of the $N$ cyclotron
orbits is expressed by the ``clustering" of the
Jacobi \\ rays: $\rho_{1} < \rho_{2} < \ldots <
\rho_{N - 1}$. Here $\rho_{1}$ is the relative
ray of the two particles with the smallest
separation, $\rho_{2}$ is the relative ray
of the nearest particle to the first two, and
so on. In the clustering approximation, comparing
the definitions (5) and (10) of the azimuthal and
the relative Jacobi angles, we have that

\begin{eqnarray}
\phi_{1j} & = & \phi_{2j} = \cdots = \phi_{j - 1 j} =
\phi_{j - 1} , \nonumber \\
j & = & 2, 3, \ldots, N,
\end{eqnarray}

\noindent and the Lagrangian (4) for $N$ anyons expressed
in Jacobi coordinates separates. The corresponding
Hamiltonian for the relative motion reads

\begin{equation}
H = \frac{1}{2m}\sum_{k = 1}^{N - 1}(p^{2}_{\rho_{k}} +
\frac{1}{\rho_{k}^{2}}(p_{\phi_{k}} + k\alpha\hbar)^{2} -
eB(p_{\phi_{k}} + k\alpha\hbar) +
\frac{1}{4}e^{2}B^{2}\rho^{2}_{k}) .
\end{equation}

\noindent The eigenvalue problem is readily solved, and
the multi-anyon wave functions $\Psi_{N}$ factorize
in the products of two-anyon wave functions

\begin{equation}
\Psi_{N} = \prod_{k=1}^{N - 1}\psi_{n_{k} , m_{k}} ,
\end{equation}

where

\begin{equation}
\psi_{n_{k} , m_{k}} = e^{ -\frac{1}{2}
\xi_{k}^{2}}L_{n_{k}}^{|m_{k} + k\alpha|}(\xi_{k}^{2})e^{
im_{k}\phi_{k}}\xi_{k}^{|m_{k} + k\alpha|} ,
\end{equation}

\noindent with $\xi_{k} = \rho_{k}\sqrt{m\omega /2\hbar}$.
The corresponding energy spectrum is the sum of two-anyon
contributions with different statistical parameters

\begin{equation}
E = \sum_{k = 1}^{N - 1}(2n_{k} + 1 + |m_{k} + k\alpha| -
m_{k} - k\alpha)\frac{\hbar\omega}{2} .
\end{equation}

The radial quantum numbers $n_{k}$ are independent
of each other and may take any positive integer value.
The angular quantum numbers $m_{k}$ can take any positive
or negative integer value, but they are however
constrained to obey the quantum analogue of the
clustering condition on the
associated classical orbits. Since $|m_{k}|$ gives
the magnitude of the corresponding Bohr-Sommerfeld orbit
$\rho_{k}$, we get the relations

\begin{equation}
|m_{k}| > |m_{k - 1}| .
\end{equation}

\noindent In addition, from the discussion of the
two-anyon problem, $m_{1}$ can take only
even integer values.

Eqn.(15) combined with rule (16) gives the spectra
drawn in Fig.3 and Fig.4 for the $3$- and the $4$-anyon
systems, as a function of $\alpha$. Neglecting
the slight non-linearity of some
of the numerical levels, we see that eqns.(15),(16)
exactly reproduce both the slopes and the correct
intercepts of the numerical eigenvalues [10].
A quick inspection of formula (15) tells
that for a generic $N$, above a certain level
there will be $N(N -1)/2$ levels with
negative slope.

The crucial ingredient in the solution of the
many-anyon problem in a magnetic field appears
to be the different topology of the
classical orbits, encircling or non-encircling each
other, which affects the angular momentum and thus the
dynamics of the system. This is not true for the
oscillator problem, where the separability of the
many-anyon problem corresponds only to classically
non-crossing trajectories, with no distinction
between encircling and non-encircling orbits. The
clustering rule (16) must then be relaxed in that
case. Thus, for the oscillator
problem the exact degeneracy is yet to be assessed.
Work is in progress in that direction [13].

If we compare the eigenfunctions (14) with
the special solutions obtained in refs. [6]-[8],
we see that the latter are included as a subset
of our solutions for extreme negative values of
the angular momentum. Furthermore, it is easy
to show in our semi-classical picture that the
zeroes of the wave functions of Wu, Chou, Dunne,
Lerda, Sciuto and Trugenberger correspond to
non-crossing classical orbits. This explains
qualitatively why we get them as a subset of
our solutions.

\vspace{1cm}

{\large \bf 5. Discussion and Conclusions}

\vspace{0.4cm}

The approach to the many-anyon problem that we
have put forward seems to have some strong
justifications from the excellent agreement
with the numerical results, from reproducing
the exact solutions previously found, and also
from the intuitive semi-classical picture.

However, much work is to be done to assess the real
connection between our solution and the complete
solution of the general many-anyon problem. In
particular, it is to be understood the role of
the new wave functions that we have derived, which
correspond to some of the linear and to all of
the slightly curved numerical eigenvalues.
This suggests the possibility of building the
exact wave functions still missing
through some perturbative or algebraic
procedure applied to our solutions [14].

In conclusion, we wish to thank Pier Alberto
Marchetti, Mario Tonin and Carlo Andrea Trugenberger
for very stimulating discussions.
\newpage
{\begin{verse} {\Large {\bf References}} \\
\vspace{1.4cm}
[1] J. M. Leinaas and J. Myrheim, Nuovo Cimento
{\bf B 37}, 1 (1977). \\
\vspace{0.8cm}
[2] F. Wilczek, Phys. Rev. Lett. {\bf 48}, 1144
(1982). \\
\vspace{0.8cm}
[3] S.M. Girvin and R. Prange, The Quantum Hall
Effect (Springer, New York, 1990). \\
\vspace{0.8cm}
[4] F. Wilczek, Fractional Statistics and Anyon
Superconductivity (World Scientific, Singapore,
1991). \\
\vspace{0.8cm}
[5] S. Forte, Rev. of Mod. Phys. {\bf 64}, 193
(1992). \\
\vspace{0.8cm}
[6] Y.S. Wu, Phys. Rev. Lett. {\bf 53}, 111
(1984). \\
\vspace{0.8cm}
[7] C. Chou, Phys. Lett. {\bf A 155}, 245 (1991). \\
\vspace{0.8cm}
[8] G. Dunne, A. Lerda, S. Sciuto, and C.A.
Trugenberger, Nucl. Phys. {\bf B 370}, 601 (1992). \\
\vspace{0.8cm}
[9] M. Sporre, J.J.M. Verbaarschot, and I. Zahed,
Phys. Rev. Lett. {\bf 67}, 1813 (1991). \\
\vspace{0.8cm}
[10] M.Sporre, J.J.M. Verbaarschot, and I. Zahed,
SUNY Preprint No. SUNY-NTG-91/40, 1991 (to be
published). \\
\vspace{0.8cm}
[11] F. Illuminati, F. Ravndal, and J.A. Ruud, Phys.
Lett. {\bf A 161}, 323 (1992). \\
\vspace{0.8cm}
[12] F. Illuminati, DFPD Preprint No. DFPD 92/TH/26,
1992 (to appear in Phys. Rev. Lett.). \\
\vspace{0.8cm}
[13] F. Illuminati, to be published. \\
\vspace{0.8cm}
[14] F. Illuminati, in preparation.
\end{verse}}
\newpage
{\Large \bf Figure Captions}

\noindent {\bf Fig.1}. Energy spectrum of two anyons in a magnetic
field, as a function of the statistical parameter.

\noindent {\bf Fig.2}. Classical orbits of two particles in a
magnetic field when the angular momentum is positive (2a),
and when it is negative (2b).

\noindent {\bf Fig.3}. Semi-classical energy spectrum of three
anyons in a magnetic field.

\noindent {\bf Fig.4}. Semi-classical energy spectrum of four
anyons in a magnetic field.
\end{document}